\documentclass[journal]{journal}
\usepackage{cite}
\ifCLASSINFOpdf
\else
\fi

\usepackage{graphicx}
\usepackage{makecell}
\usepackage{harveyballs}
\usepackage{siunitx}
\usepackage[disable]{todonotes}
\usepackage[nogroupskip,acronyms,nopostdot,style=long,nonumberlist,toc]{glossaries}
\usepackage{hyperref}
\usepackage{tikz}
\usetikzlibrary{arrows.meta,plotmarks,calc}
\usepackage{pgfplots}
\usepgfplotslibrary{groupplots,units,external,fillbetween}
\pgfplotsset{compat=1.15}
\usepackage{pdflscape}
\usepackage{subcaption}
\usepackage{pdfcomment}
\usepackage{booktabs}

\newcommand*\circled[1]{\tikz[baseline=(char.base)]{\node[shape=circle,draw,inner sep=1pt] (char) {\footnotesize #1};}}
\newcommand{\quotes}[1]{``#1''}

\newlength{\thickarrayrulewidth}
\newacronym{ac}{AC}{Alternating Current}
\newacronym{au}{AU}{African Union}
\newacronym{bev}{BEV}{Battery Electric Vehicle}
\newacronym{bmz}{BMZ}{German Federal Ministry for Economic Cooperation and Development}
\newacronym[\glslongpluralkey={customer-relevant properties},\glsshortpluralkey={CRPs}]{crp}{CRP}{customer-relevant property}
\newacronym{cp}{CP}{Concept Phase}
\newacronym{dc}{DC}{Direct Current}
\newacronym{dsr}{DSR}{Design Science Research}
\newacronym{dp}{DP}{Define Phase}
\newacronym{eu}{EU}{European Union}
\newacronym[\glslongpluralkey={Focus Group Discussions},\glsshortpluralkey={FGDs}]{fgd}{FGD}{Focus Group Discussion}
\newacronym[\glslongpluralkey={Fuel Cell Electric Vehicles},\glsshortpluralkey={FCEVs}]{fcev}{FCEV}{Fuel Cell Electric Vehicle}
\newacronym{ftm}{FTM}{Institute of Automotive Technology}
\newacronym{gdp}{GDP}{Gross Domestic Product}
\newacronym{giz}{GIZ}{German Cooperation}
\newacronym[\glslongpluralkey={Hybrid Electric Vehicles},\glsshortpluralkey={HEVs}]{hev}{HEV}{Hybrid Electric Vehicle}
\newacronym{iaa}{IAA}{International Motor Show}
\newacronym{ilo}{ILO}{International Labor Organization}
\newacronym[\glslongpluralkey={Intermediate Means of Transport},\glsshortpluralkey={IMTs}]{imt}{IMT}{Intermediate Means of Transport}
\newacronym{ip}{IP}{Ideate Phase}
\newacronym{irap}{IRAP}{Integrated Rural Accessibility Planning}
\newacronym{lpl}{LPL}{Local Project Leader}
\newacronym[\glslongpluralkey={Primary Cooperatives},\glsshortpluralkey={PCs}]{pc}{PC}{Primary Cooperative}
\newacronym[\glslongpluralkey={Points of Interest},\glsshortpluralkey={POIs}]{poi}{POI}{Point of Interest}
\newacronym{pra}{PRA}{Participatory Rural Appraisal}
\newacronym{pp}{PP}{Prototype Phase}
\newacronym{ssa}{SSA}{Sub-Saharan Africa}
\newacronym{ssatp}{SSATP}{Sub-Saharan Africa Transport Policy Program}
\newacronym{tp}{TP}{Test \& Validation Phase}%Half spaces not allowed in hyperref together with pdfcomment
\newacronym{tum}{TUM}{Technical University of Munich}
\newacronym{ukaid}{UKAID}{British Cooperation}
\newacronym{up}{UP}{Understand \& Empathize Phase}%Half spaces not allowed in hyperref together with pdfcomment
\newacronym{un}{UN}{United Nations}

\usepackage[none]{hyphenat}
\usepackage[justification=centering]{caption}	
\usepackage[labelsep=space]{caption}		
\usepackage[font=footnotesize]{caption}
\begin{document}

\title{Enhancing Rural Agricultural Value Chains through Electric Mobility Services in Ethiopia}
\author{Clemens~Pizzinini, Philipp~Rosner, David~Ziegler, Markus~Lienkamp
\thanks{Clemens~Pizzinini, Philipp~Rosner, David~Ziegler, and Markus~Lienkamp are affiliated with the Institute of Automotive Technology, Technical University of Munich, Boltzmannstrasse 15, 85748 Garching, Germany (e-mail: clemens.pizzinini@tum.de).}% <-this % stops a space
}


\markboth{Journal of \LaTeX\ Class Files,~Vol.~6, No.~1, January~2007}%
{Shell \MakeLowercase{\textit{et al.}}: Bare Demo of IEEEtran.cls for Journals}



\maketitle
\thispagestyle{empty}

\begin{abstract}
Transportation is a constitutional part of most supply and value chains in modern economies. Smallholder farmers in rural Ethiopia face severe challenges along their supply and value chains. In particular, suitable, affordable, and available transport services are in high demand. To develop context-specific technical solutions, a problem-to-solution methodology based on the interaction with technology is developed. With this approach, we fill the gap between proven transportation assessment frameworks and general user-centered techniques. Central to our approach is an electric test vehicle that is implemented in rural supply and value chains for research, development, and testing. Based on our objective and the derived  methodological requirements, a set of existing methods is selected. Local partners are integrated in an organizational framework that executes major parts of this research endeavour in Arsi Zone, Oromia Region, Ethiopia. 
\end{abstract}

\begin{IEEEkeywords}
Agricultural value chain, participatory methods, agile methods, sub-Saharan Africa, Ethiopia, electric vehicle, transport service.
\end{IEEEkeywords}

% \fi
\IEEEpeerreviewmaketitle

\section{Introduction}
\IEEEPARstart{T}{he} value chain is one of the fundamental frameworks in modern economics. Rural agricultural value chains in \gls{ssa} and their sustainable development are an integral part of many development projects. A value chain represents a firm's activities that create economic value by transforming resources into products or services \cite{Porter.2001}. Economic value is defined as the price a customer is willing to pay. A firm is profitable if the value created exceeds the cost of producing the product or delivering the service \cite{Porter.2001}. The set of decisions on what activities to specialize in and which ones to outsource is part of every firm's strategy. The economic benefits of successfully balancing this trade-off lead to increasing effectiveness and generating value \cite{Quinn.1994}. Once a firm decides to outsource a number of value-creating activities to specialize, it enters into with other firms that supply or are supplied by the firm. This network of value-creating participants is known as supply chain \cite{LaLonde.1994}. An activity that firms in modern economies frequently outsource is transportation. Labor cost reduction, specialization, and asset reduction are the main reasons to access transportation as a service \cite{Bardi.1991}.

Agricultural value chains in Ethiopia account for \SI{36}{\percent} of the country's \gls{gdp}. More than \SI{70}{\percent} of the population is engaged in agriculture \cite{Degu.2019}. About \SI{95}{\percent} of the agricultural output originates from smallholder farmers \cite{Abebaw.2013} with an average plot size of \SI{0.85}{\hectare} \cite{Abay.2020}. Ethiopian farmers in particular have been experiencing declining plot sizes with ongoing population growth \cite{Abay.2020}. Ceteris paribus, this results in less marketable produce which subsequently reduces the ability to make decisions about specialization and the configuration of the value and supply chain.

The research approach developed in this paper analyzes the problem from a supply point of view. We aim to answer how we design vehicle-based services that enhance smallholder farmer's value chains. We consider modern electric vehicles as mobile platforms for innovation. The to-be-developed research approach shall allow for a holistic investigation of rural supply and value chains in Ethiopia to develop a highly relevant service.

\subsection{Transportation: A Wicked Problem}
"Rural people need transport that is affordable, predictable and dependable, timely, safe and secure and that can carry people’s goods and (when required) their supporting persons" \cite{Starkey.2016}. Based on this information, we define transportation service quality perceived by users along three main \glspl{crp}: vehicle type, service availability and affordability. From the vehicle operator perspective, however, the offered quality of each  \pdftooltip{\gls{crp}}{\glsentrydesc{crp}} is related to a set of costs. Wittenbrink \cite{Wittenbrink.2014} defines these costs as variable costs (energy costs, maintenance, etc.), staff costs (driver), fixed costs (acquisition cost, insurance, etc.), and overheads (management). The most pressing challenges rural transport operators face in  \gls{ssa} generally are increasing costs due to remote and scattered communities \cite{Sieber.2009,Poulton.2010}, poor road quality \cite{Dennis.2017,Ellis.1998} (especially during the rainy season \cite{Hine.2014}) and types of vehicles (passenger cars for heavy-duty application) \cite{Dennis.2017}. Low income across their customer base \cite{Bryceson.2008} further increases the pressure to operate cost-effectively and to overcome vehicle underutilization \cite{Ellis.1998}. These deterring circumstances decrease the number of transport operators, which leads to low competition and increasing prices for users \cite{Ellis.1998}. Rural agricultural supply chains, therefore suffer from deficits in all three \pdftooltip{\glspl{crp}}{\glsentrydesc{crp}}. Consequently, the farmers' ability to take the introduced specialization and outsourcing decision is constrained.
\subsection{Methodologies to Assess and Develop Rural Transportation Services}
There are several publications on rural transport services in SSA. These include works of researchers affiliated with a research institution and publications executed or funded by international (development) organizations such as the World Bank, \gls{ukaid}, \gls{giz}, European Union, United Nations, and the African Union. A range of publications survey and analyze primary data to understand a well-defined detail within the supply and value chains of rural farmers. For example, this might be the price \,\&\, the cost structure of transportation in a specific area \cite{Njenga.2015,Njenga.2014,Dennis.1998,Dennis.2017,Ellis.1998}, available modes of (intermediate) transport, or travel times to the market \cite{Njenga.2014}. Additionally, supply and value chains of specific highly valuable agricultural products like tomatoes are surveyed \cite{Sieber.2019}. Besides surveys, traffic counting is used to quantify current mobility behavior at specified points \cite{Njenga.2015,Njenga.2014,Dennis.1998}. Few publications talk about technological interventions that are based on previously identified requirements \cite{Dennis.2017,Soltes.2018d,Starkey.2019}. 

Three approaches are highlighted because of their relevance to this domain of research. As part of the World Bank's \gls{ssatp}, Starkey \cite{Starkey.2007b} introduces a rapid appraisal survey design for assessing transportation accessibility in a defined geographic area. Results from semi-structured interviews with transport operators, regulators, users, and support services are triangulated with traffic counts and other primary data within a relatively short period of time. Commissioned by the International Labour Organization, Donnges et\,al. \cite{Donnges.2003} develop a planning process called Integrated Rural Accessibility Planning for government-level users including accessibility mapping, community participation workshops, and project identification. Funded by the UK Department for International Development, Bryceson et\,al. \cite{Bryceson.2003} utilize the Livelihood Approach to identify the mobility and accessibility needs of target groups in Zimbabwe and Uganda. In two phases, the study uses Focus Group Discussion, key informant interviews, and household questionnaires to compare the accessibility, mode of transport and mobility needs of different income groups. The remaining publications are review reports based on secondary survey data to derive policy and project recommendations on a macro level \cite{Sieber.2009,Starkey.2019,Banjo.2012,Booth.2000}. 

Neither of the relevant identified methodologies include a holistic problem-to-solution development, nor is a majority focused on the individual rural households' demands \cite{Banjo.2012}. Most publications conclude with systematic recommendations to regulatory stakeholders. This is indubitably important but omits the necessary validation step to link cause and effect. Literature also suggests that many investments into the transport sector have missed the preceding intended impact due to emphasizing the wrong problem scale or a sole focus on infrastructure investments without the provision of appropriate services \cite{Booth.2000}. Apart from acknowledging the rapid increase in cellphone penetration \cite{Minten.2020} and conventional motorcycles as a surveyed item that might be available to the user, only one publication deals with the potential of modern technologies like electric vehicles within rural supply and value chains \cite{Soltes.2018d}.  
\subsection{Research Question}
It can be seen from the literature that there is a need for an approach that combines assessment frameworks with methods of general problem-solving based on the application of modern technologies. Therefore, this paper aims to develop a research approach to understand challenges and develop solutions to address the three \pdftooltip{\glspl{crp}}{\glsentrydesc{crp}} of transportation services.
\section{Developing a Novel Research Approach}
"Research approaches are plans and the procedures for research that span the steps from broad assumptions to detailed methods of data collection, analysis, and interpretation" \cite{Creswell.2017}

According to Creswell et\,al. \cite{Creswell.2017}, a research approach is framed by a philosophical worldview, the research design, and its applied methods. Furthermore, the nature of the research problem, the researcher's personal experiences, and the audience of any result need to be considered preliminarily. As engineers, we follow a pragmatic problem-solving approach. By focusing on the identified problem rather than on particular methodologies and their ontological nature, this paper aims to develop a requirement-driven research approach. The deployed methods are selected according to their expected contribution to the problem. The following paragraphs first introduce Design Science Research as the underlying school of thought for such a research approach. A set of context and project-specific requirements is derived before benchmarking existing research frameworks and methodologies against them.
\subsection{Requirements for Design Science Research}
"Historically and traditionally, it has been the task of science disciplines to teach about natural things: how they are and how they work. It has been the task of engineering schools to teach about artificial things: how to make artifacts that have desired properties and how to design them" \cite{Simon.1988}.

Excluding direct carriage by humans or animals, transportation is an activity residing on the utilization of technical machinery like cars, motorcycles, animal carts, or bicycles. Consequently, a problem-to-solutions approach (research, development, and testing) includes either the design of new or the optimization of existing artifacts. Peffers et\,al. \cite{Peffers.2007} summarize a set of practical rules such a research endeavor shall comply with: The created artifact must address the problem, be relevant, and its utility, quality, and efficacy should be evaluated. Furthermore, the development process must consist of proven methods and its results being communicated to relevant stakeholders. 

Several general approaches facilitate complex human-centered problem-solving. The most prominent one is called \quotes{Design Thinking}. Within a creative five-phase approach, an interdisciplinary team works towards designing a specifically suitable solution for defined target groups \cite{Meinel.2016}. In the Understand \& Empathize Phase, methods from ethnology are applied to investigate the living context of the target group to derive a well-defined problem and target group definition during the Define Phase. Expanding from that, the following Ideate Phase, Prototype Phase, and Test \& Validation Phase iteratively generate solutions \cite{Meinel.2016}. Each of these is tested frequently, collecting feedback from the target group to increase acceptance and ensure quick and coordinated solution development  \cite{Meinel.2016}.

Participatory Rural Appraisal is an overarching description for an extensive set of methods to enable local people to share, enhance, and analyze their knowledge to plan and act for a better standard of living \cite{Chambers.1994}. Participatory Rural Appraisal highlights the importance of secondary sources, verbal interaction, and observations. Its origins can be found in research on farming systems and agro-economic analysis \cite{Chambers.1994}.

A more recent method to foster user-centered research and co-creation is Living Labs. The basic concept behind them is the involvement of the entire value chain from technology suppliers, content and technology providers, and the end customer or consumer in a setting that is investigated \cite{Eriksson.2006}. Within this real-world laboratory, it is possible to grasp the value and implications specific technological interventions offer. The collaboration of different actors within the Living Labs is called open-innovation networks \cite{Leminen.2012} that can enhance user acceptance, integrate existing knowledge, and increase sustainability.

These three general approaches inspired us to consider the user as an active member in the research, development, and testing by increasing interactions between the user and technology. Nevertheless, as the objective, technology, and users of our endeavor are highly contextual, a research approach needs to be specified accordingly. 
\begin{table}[!htbp]
\scriptsize
\centering
\caption{Assessment of existing methods }
\label{tab:RatingExistMeth}
\begin{tabular}{rccccccc}
\toprule
& \multicolumn{7}{c}{Objective No.}\\
Method  & \circled{1} & \circled{2} & \circled{3} & \circled{4} & \circled{5} & \circled{6} & \circled{7}\\
\toprule
Starkey et\,al. \cite{Starkey.2007b} & \harveyBallHalf & \harveyBallHalf & \harveyBallHalf & \harveyBallQuarter & \harveyBallHalf & \harveyBallHalf & \harveyBallQuarter\\
Donnges et\,al. \cite{Donnges.2003}& \harveyBallQuarter & \harveyBallFull & \harveyBallHalf & \harveyBallFull &  \harveyBallFull & \harveyBallFull & \harveyBallFull\\
Vajjhala et\,al. \cite{Vajjhala.2010}& \harveyBallNone & \harveyBallHalf & \harveyBallNone & \harveyBallQuarter & \harveyBallFull & \harveyBallQuarter & \harveyBallQuarter\\
Njenga et\,al. \cite{Njenga.2014}& \harveyBallHalf & \harveyBallHalf & \harveyBallHalf & \harveyBallNone & \harveyBallQuarter & \harveyBallQuarter & \harveyBallNone\\
Siebert et\,al. \cite{Sieber.2019}& \harveyBallHalf & \harveyBallHalf & \harveyBallHalf & \harveyBallNone & \harveyBallQuarter  & \harveyBallQuarter & \harveyBallQuarter\\
Bryceson et\,al. \cite{Bryceson.2003}& \harveyBallHalf & \harveyBallHalf & \harveyBallHalf & \harveyBallNone & \harveyBallQuarter  & \harveyBallQuarter & \harveyBallNone\\
Dennis \cite{Dennis.1998}& \harveyBallHalf & \harveyBallHalf & \harveyBallQuarter & \harveyBallNone & \harveyBallQuarter  & \harveyBallQuarter & \harveyBallNone\\
Ellis et\,al. \cite{Ellis.1998}& \harveyBallThreeQuarter & \harveyBallThreeQuarter & \harveyBallHalf & \harveyBallThreeQuarter & \harveyBallHalf  & \harveyBallHalf & \harveyBallNone\\
\bottomrule
\end{tabular}
\end{table}

\subsection{Context and Project-specific Requirements}
Differences between the context in which a technical solution is developed, and the one in which it is implemented are detrimental to its properties and cannot be neglected. To translate the general practical rules imposed by Design Science Research and the fundamental ideas from general approaches into more context-specific requirements, the objective of this research approach is further specified: The main objective is to facilitate a user-centered research and development process accompanied by continuous data gathering and analysis. Focusing on the individual farmers' challenges in the agricultural value and supply chain in Ethiopia, a sustainable transport service model needs to be developed specifically for and with this target demographic. The key performance indicators are the three \pdftooltip{\glspl{crp}}{\glsentrydesc{crp}} affordability, availability, and vehicle type. The result shall allow for a quantified assessment of electric vehicles' transformation potential within the agricultural sector in Ethiopia. Based on this main objective, a set of requirements is derived (Table \ref{tab:ResApprReq}) to further assess supplementary methods.
\begin{table}[!htb]
\scriptsize
\centering
\caption{Requirements for research approach}
\label{tab:ResApprReq}
\begin{tabular}{p{0.3cm}p{1.0cm}p{2.0cm}p{3.5cm}}
\toprule
No. & CRP & Outcome & Research approach requirement\\
\toprule
\circled{1} & Cost & Sustainable business model & \makecell[l]{Cost \& revenue analysis\\Value \& supply chain focus\\Vehicle-based service development}\\
\midrule
\circled{2} & Availability & Efficient operation model & \makecell[l]{Transport demand analysis\\Energy demand analysis}\\
\midrule
\circled{3} & Vehicle & Suitable vehicle concept & User-vehicle interaction\\
\midrule
\circled{4} & \makecell[l]{Cost\\Availability} & Feasible system integration & \makecell[l]{Energy access assessment\\Existing fleet assessment\\Road assessment}\\
\midrule
\circled{5} & \makecell[l]{Cost\\Availability\\Vehicle} & High user acceptance & \makecell[l]{Co-development\\Inclusiveness\\Capacity building}\\
\midrule
\circled{6} & General & High stakeholder acceptance & \makecell[l]{Integration of stakeholders\\Transfer of ownership\\Leveraging existing structures\\Capacity building}\\
\midrule
\circled{7} & General & Validated solution & \makecell[l]{Contextual parameter definition\\Longitudinal data collection\\Data triangulation\\Real world testing}\\
\bottomrule
\end{tabular}
\end{table}
\subsection{Assessment of Existing Methods}
The identified transport-related research methods are assessed in Table \ref{tab:RatingExistMeth} according to the outcome set out in Table \ref{tab:ResApprReq}. The general problem-solving methods are excluded from the assessment as their methodology is highly generalized. Nevertheless, their introduction and understanding are essential as they represent the paradigmatic foundation of the proposed research approach. Harvey Balls are used to rate the fit on the following scale: \harveyBallNone\,-\,not considered, \harveyBallQuarter\,-\,proposed,  \harveyBallHalf\,-\,analysed, \harveyBallThreeQuarter\,-\,developed, and \harveyBallFull\,-\,tested.   

\subsection{Generalization of Results for Potential Evaluation}
As the sciences’ objective is generalizability, the to-be-developed research approach must address its ability to produce valid solutions for similar problems in changing circumstances (transferability) and valid solutions for -to limited extent-related problems. Transferability starts with the stated research and development objectives that apply to similar regions with shortcomings along agricultural supply and value chains in \pdftooltip{\gls{ssa}}{\glsentrydesc{ssa}}. This further applies to the derived methodological requirements. The final selection of the implemented methods depends on the country and regionally-specific properties like the availability of traffic data and needs to be evaluated beforehand. The question of whether the produced results can be scaled on a regional or even national level in Ethiopia poses the fundamental question of the general resides in particular. In an economic and therefore social science setting,  Hill et\,al. \cite{Hill.1999} confirm this as a widely held belief in this domain. Its validity for engineering endeavors is at least to be questioned.
\section{The aCar Mobility Research Approach}
The proposed research approach is the result of an analysis of specific transport-oriented approaches focused on the application in and for \pdftooltip{\gls{ssa}}{\glsentrydesc{ssa}} and general complex sociotechnical problem-solving methods. It is a mixed-method approach within a scenario similar to a Living Lab, involving all relevant stakeholders over a period of two years. The following sections give a detailed overview of the research phases, the methodological foundation, specific interventions, and the technical artifacts.

\subsection{Research Phases}
\begin{figure*}[!htbp]
\includegraphics[width=\textwidth]{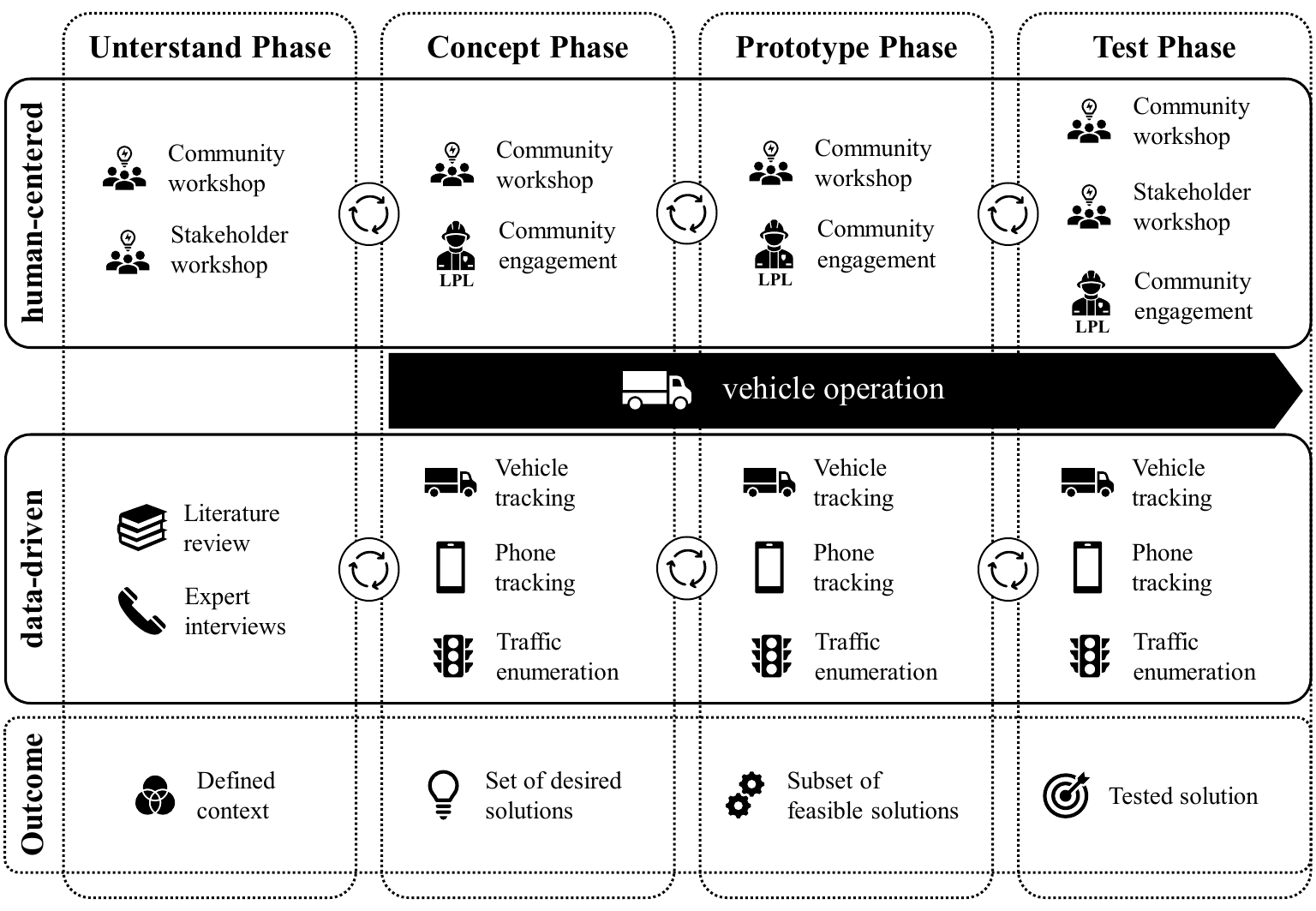}
%\missingfigure{}
\caption{Matrix view of chronological and methodological research approach}
\label{fig:ResAppr}
\end{figure*}

The process is based on Design Thinking and is divided into four distinct research phases. These are initiated sequentially but allow for iterations if necessary. Since the target user group is predefined as rural farmers, Design Thinking's Definition Phase is omitted. Starting with the \gls{up}, the contextual system surrounding rural supply and value chains in the target area is described and parameterized. The resulting system design is then transferred to the renamed \gls{cp} for the co-development of an extensive set of desired solution concepts. The prioritized concepts are then translated into functional but very simple prototypes within the \gls{pp}. A subset of feasible prototypes is tested throughout the \gls{tp}. The procedure takes up to two years but should not be shorter than one year to safeguard user and stakeholder acceptance and trust in the developed technology. Figure \ref{fig:ResAppr} uses a matrix view. While the four phases' desired outcomes are arranged horizontally, the vertical axis separates the methodological categories of human-centered (qualitative) and data-driven (quantitative) methods. Each field within this matrix shows the main methods used. These will be further detailed in section \ref{sec:MixedMethods}.
\subsection{Mixed-method Approach\,\&\,Triangulation}
\label{sec:MixedMethods}
In all four research phases, various methods based on the requirements derived in Table \ref{tab:ResApprReq} are applied. A majority of them are based on the presented literature, but additional methods specific to this research approach are proposed (Table \ref{tab:ApplMeth}). Combining quantitative and qualitative methods to provide a more comprehensive image, this research approach can be classified as a \quotes{Convergent Mixed Method} approach \cite{Creswell.2017}. Triangulating results from different methods for the same phenomena, supports the analysis process \cite{Hill.1999}, and eases plausibilization. In the following, the most important ones are presented.
\subsubsection{Community\,\&\,Stakeholder Workshops}
Within each research phase, a community workshop involving the target group of farmers is hosted, enabling a continued involvement and co-creation process. The workshop duration is 2\,-\,3 days to cater for initial group forming activities necessary in a larger group of about 20 participants. The participant selection is the same for all four workshops to facilitate analysis over the project duration. A workshop moderator facilitates the process in the local language. Each workshop's topics are based on those of the respective research phase. During the \pdftooltip{\gls{up}}{\glsentrydesc{up}}, the participants work to e.\,g. identify challenges and systematic problems along their agricultural supply and value chains, whereas in the \pdftooltip{\gls{cp}}{\glsentrydesc{cp}}, efforts are focused on theoretical solution concepts.
Both at the beginning and at the end of the project, a stakeholder workshop is held. Representatives from local authorities, transport operators, and international organizations active in the agricultural sector participate in a single-day event to discuss research interventions and outcomes. This workshop is particularly important to create awareness amongst stakeholders and gain support for activities.
\subsubsection{Vehicle Operation}
Throughout the project duration, an electric vehicle is continuously operated as a test platform. Its selection process is described in section  \ref{sec:VehicleSelec}. Initially, the vehicle offers basic on-demand transportation to participating communities. At this point, valuable technical data about road quality, energy consumption, charging, and grid reliability is gathered. Additionally, the users are introduced to the technology to build trust and acceptance. In later research phases, the vehicle operation scheme is modified as a test platform for solutions.
\subsubsection{Local Research Partnership}
To increase user and stakeholder acceptance, incorporate local knowledge, and build local capacity, a local research partner is selected to execute major parts of the research approach. Local research staff is responsible for gathering data, supervising operations, hosting workshops, and conducting surveys. A regular report (in our case following a biweekly schedule) is gathered from other involved persons and forwarded to the responsible research team in Germany.
\subsubsection{Literature Review}
In a pre-study, relevant literature on rural transportation, agricultural supply\,\&\,value chains, mobility analysis, electrification (on and off-grid), and vehicle concepts with focus on \pdftooltip{\gls{ssa}}{\glsentrydesc{ssa}} is identified and screened. An extensive set of 50 influential contextual parameters is derived. These parameters are evaluated through the mentioned workshops and semi-structured interviews, as well as the quantitative data gathering.
\subsubsection{Semi-structured Interviews}
Since the literature on contextual parameters in the specific intervention area is scarce, semi-structured interviews with users, operators, and other stakeholders allow for novel perspectives. Results from the pre-study are translated into a general framework of themes. Newly addressed issues are first tested against literature before being included in the framework of themes for quali- and quantification.
\begin{table}[!htbp]
\scriptsize
\centering
\caption{Applied methods based on research approach requirement}
\label{tab:ApplMeth}
\begin{tabular}{p{0.5cm}p{2.7cm}p{3.0cm}p{1.0cm}}
\toprule
Obj. & Research approach requirement & Method & Phase(s)\\
\toprule
\circled{1} & Cost \& revenue analysis & \makecell[l]{User interviews\\Operator interviews\\ Literature analysis\\Continuous vehicle operation} & \makecell[l]{\pdftooltip{\gls{up}}{\glsentrydesc{up}}\\\pdftooltip{\gls{up}}{\glsentrydesc{up}}\\\pdftooltip{\gls{up}}{\glsentrydesc{up}}\\All}\\
\midrule
\circled{1} & Value \& supply chain focus & \makecell[l]{Stakeholder workshop interviews\\Community workshops\\ Literature analysis} & \makecell[l]{\pdftooltip{\gls{up}}{\glsentrydesc{up}}\\All\\\pdftooltip{\gls{up}}{\glsentrydesc{up}}}\\
\midrule
\circled{1} & \makecell[l]{Vehicle-based\\service development} & \makecell[l]{Stakeholder workshops\\Community workshops\\ Literature analysis} & \makecell[l]{\pdftooltip{\gls{up}}{\glsentrydesc{up}}\\\pdftooltip{\gls{cp}}{\glsentrydesc{cp}}\\\pdftooltip{\gls{up}}{\glsentrydesc{up}}}\\
\midrule
\circled{2}& \makecell[l]{Transport demand analysis\\Energy demand analysis} & \makecell[l]{GPS mobility tracking\\Traffic flow enumeration\\ Literature analysis} & \makecell[l]{All\\\pdftooltip{\gls{cp}}{\glsentrydesc{cp}}\\\pdftooltip{\gls{up}}{\glsentrydesc{up}}}\\
\midrule
\circled{3} & User-vehicle interaction & Continuous vehicle operations & All\\
\midrule
\circled{4} & Energy access assessment & \makecell[l]{Stakeholder workshops\\Community workshops\\ Literature analysis} & \makecell[l]{\pdftooltip{\gls{up}}{\glsentrydesc{up}}\\\pdftooltip{\gls{up}}{\glsentrydesc{up}}\\\pdftooltip{\gls{up}}{\glsentrydesc{up}}}\\
\midrule
\circled{4} & Existing fleet assessment & \makecell[l]{Community workshops\\ Literature analysis\\Traffic flow enumeration} & \makecell[l]{\pdftooltip{\gls{up}}{\glsentrydesc{up}}\\\pdftooltip{\gls{up}}{\glsentrydesc{up}}\\\pdftooltip{\gls{cp}}{\glsentrydesc{cp}}}\\
\midrule
\circled{4} & Road assessment & \makecell[l]{Continuous vehicle operations\\Aerial image analysis} & \makecell[l]{All\\\pdftooltip{\gls{cp}}{\glsentrydesc{cp}}}\\
\midrule
\circled{5} & Co-development & \makecell[l]{Community workshops\\Continuous vehicle operations} & \makecell[l]{All\\All}\\
\midrule
\circled{5} & Inclusiveness & \makecell[l]{\SI{>40}{\percent} females in workshops\\Female-specific value chains} & \makecell[l]{All\\All}\\
\midrule
\circled{5} & Capacity Building & \makecell[l]{Driver training\\Technician training\\Community workshops} & \makecell[l]{\pdftooltip{\gls{cp}}{\glsentrydesc{cp}}\\\gls{cp}\\All}\\
\midrule
\circled{6} & Integration of stakeholders & \makecell[l]{Stakeholder workshops\\Stakeholder interviews} & \makecell[l]{\pdftooltip{\gls{up}}{\glsentrydesc{up}}\,\&\,\pdftooltip{\gls{tp}}{\glsentrydesc{tp}}\\\pdftooltip{\gls{up}}{\glsentrydesc{up}}}\\
\midrule
\circled{6} & Transfer of ownership & \makecell[l]{Partnership with local university\\Local staffing\\Farmers' union involvement} & \makecell[l]{All\\All\\All}\\
\midrule
\circled{6} & Leveraging existing structures & Farmers' union involvement & All\\
\midrule
\circled{7} & Contextual parameter analysis & \makecell[l]{Literature review\\Expert interviews} & \makecell[l]{\pdftooltip{\gls{up}}{\glsentrydesc{up}}\\\pdftooltip{\gls{up}}{\glsentrydesc{up}}}\\
\midrule
\circled{7} & Longitudinal data collection & \makecell[l]{Continuous vehicle operation\\Traffic enumeration} & \makecell[l]{All\\\pdftooltip{\gls{cp}}{\glsentrydesc{cp}}}\\
\midrule
\circled{7} & Data triangulation & \makecell[l]{Rolling data analysis\\Mixed-Methods} & \makecell[l]{All\\All}\\
\midrule
\circled{7} & Real world testing & \makecell[l]{Continuous vehicle operation\\Community workshops} & \makecell[l]{All\\All}\\
\bottomrule
\end{tabular}
\end{table}
\subsection{Involved Partners}
To facilitate the research approach and methods, a multi-level structure of different research partners is developed. A local research partner (university) oversees all data-gathering activities, hosts community workshops, and maintains the vehicle's operation. The university staff directly communicates with a farmers' union, representing several \glspl{pc}. The \pdftooltip{\glspl{pc}}{\glsentrydesc{pc}} themselves are split up into grain and seed-producing cooperatives. In Ethiopia, most individual farmers are organized in cooperatives to receive agricultural inputs and services \cite{Mojo.2018,Getnet.2012}. The vehicle is stationed at the union's compound and serves a selection of two to three \pdftooltip{\glspl{pc}}{\glsentrydesc{pc}} within a radius of \SI{20}{\kilo\meter}. The participating farming communities are exclusively members of the cooperatives. 

A \gls{lpl} is employed by the union. He or she oversees daily operations, executes the plan set out by the universities, and maintains engagement with the community members, being their direct access point to the project. He or she issues a regular (in this case biweekly) report containing the driving routes, customers served, goods transported, and faced issues. During the community workshops, the \pdftooltip{\gls{lpl}}{\glsentrydesc{lpl}} acts as a spokesman for the community. 
\begin{figure}[!htbp]
\centering
\includegraphics[width=\columnwidth]{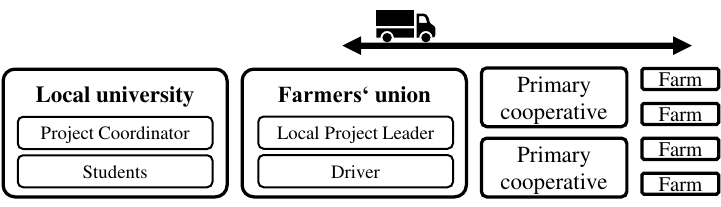}
%\missingfigure{}
\captionsetup{justification=centering}
\caption{Overview of the partner structure, including the respective roles within the project}
\label{fig:StakeStruct}
\end{figure}
\subsection{Test Vehicle}
\label{sec:VehicleSelec}
The employed vehicle type is one of the three transport services \pdftooltip{\glspl{crp}}{\glsentrydesc{crp}}. It, therefore is of significant importance to the success of the proposed research approach. Apart from being an electric vehicle and therefore accurately modeling their possible future implementation, several other requirements arise from the \pdftooltip{\gls{ssa}}{\glsentrydesc{ssa}} test setting, listed in Table \ref{tab:TestVehReq}. In short, a simple and smart vehicle is required \cite{Soltes.2018d}, with a smart vehicle being one that is well-suited to the context it is being implemented into. While multiple concepts for electrified drive trains exist, the most common being the \gls{hev}, \gls{bev} and \gls{fcev}, only the \pdftooltip{\gls{bev}}{\glsentrydesc{bev}} can operate without harmful gaseous and acoustic emissions, alleviating local as well as global pollution problems while simultaneously being compatible to existing energy supply infrastructure in \pdftooltip{\gls{ssa}}{\glsentrydesc{ssa}}.

\begin{figure}[!htbp]
\centering
\includegraphics[width=\columnwidth]{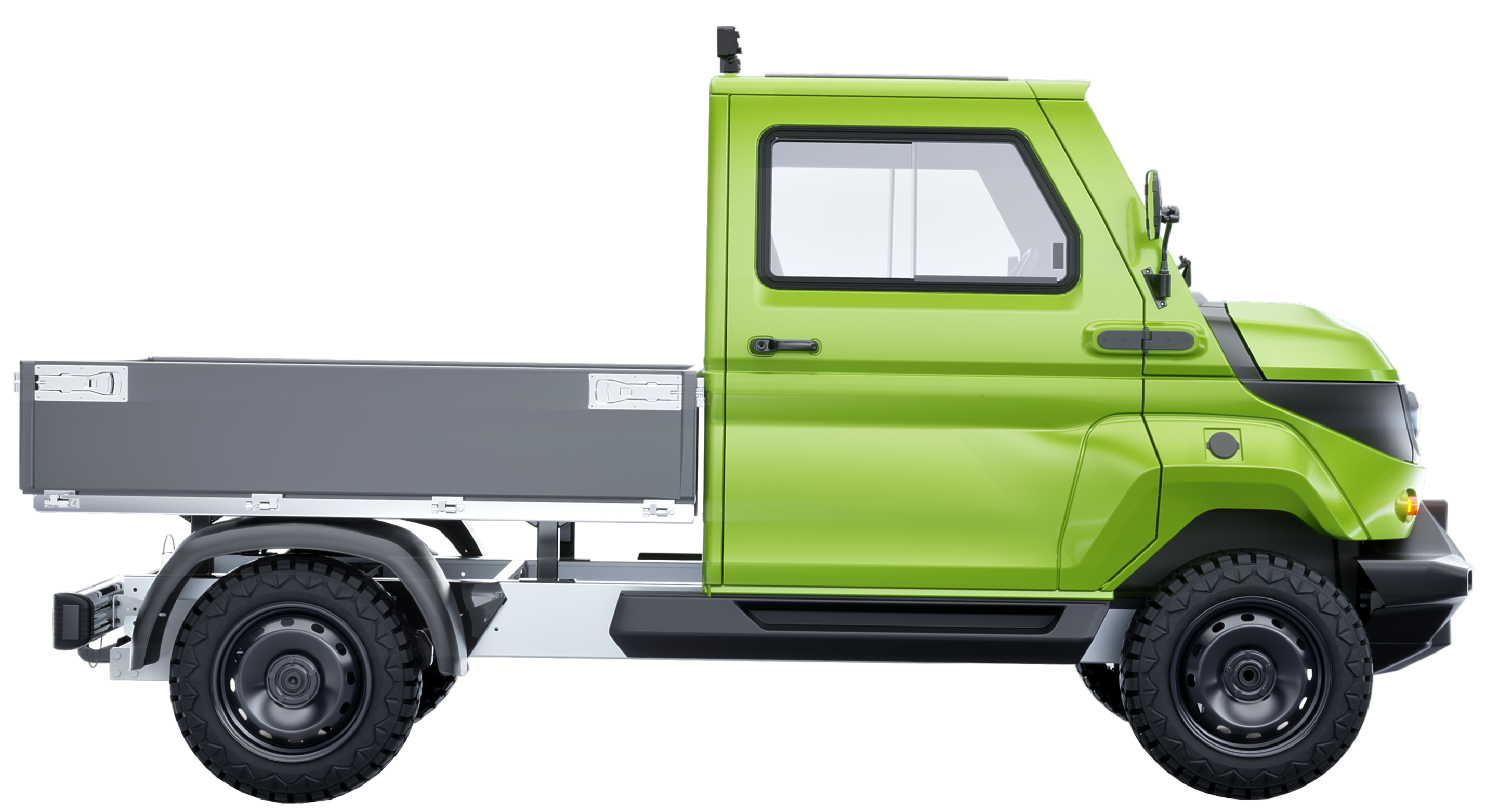}
%\missingfigure{}
\captionsetup{justification=centering}
\caption{EVUM Motors aCar \cite{EVUMMotorsGmbH.2019}}
\label{fig:acar}
\end{figure}

The requirements listed strongly differ from those typically for passenger vehicles in highly industrialized countries, like high speed, connectivity, and automated functions. Therefore, such vehicles are excluded, and an alternative specifically designed for similar requirements is sought. In 2014, the \pdftooltip{\glsentryfull{ftm}}{\glsentrydesc{ftm}} at \pdftooltip{\gls{tum}}{\glsentrydesc{tum}} initiated a research and development project for an electric vehicle concept for rural \pdftooltip{\gls{ssa}}{\glsentrydesc{ssa}}, called the \quotes{aCar} \cite{Soltes.2018d}. In 2017, a fully functioning prototype was exhibited at the \gls{iaa} in Frankfurt am Main, Germany, while another was tested in Ghana. In 2020, EVUM Motors GmbH started series production of a further refined version in Germany. Due to its bespoke concept for use in rural \gls{ssa}, it fulfills the listed requirements sufficiently as shown with Harvey Balls in the rightmost column of Table \ref{tab:TestVehReq} and will therefore be used as this research concept's test vehicle. Its shortcomings are mainly based on the currently high purchase price compared to conventionally fueled vehicles, which will alleviate with rising production numbers, and the currently inexisting spare parts supply in \pdftooltip{\gls{ssa}}{\glsentrydesc{ssa}}, which requires importing the needed parts from Germany within the project. Competitors in the light electric truck sector are usually designed for \pdftooltip{\gls{eu}}{\glsentrydesc{eu}} municipal\,\&\,commercial customers, therefore following different priorities and not being suitable for the target scenario. Fig.\,\ref{fig:acar} shows the selected test vehicle in its current form.
\begin{table}[!htbp]
\scriptsize
\centering
\caption{Test vehicle requirements and fulfillment by the selected vehicle}
\label{tab:TestVehReq}
\begin{tabular}{l@{\hskip 0.1cm}lc}
\toprule
Operational aspect & Vehicle requirements & aCar\\
\toprule
Affordability & \makecell[l]{Low initial purchase price\\Low variable cost\\Longevity} & \harveyBallQuarter\\
\midrule
\makecell[l]{Locally emission-free operation\\Independence from fuel oil supply} & Exclusion of \glspl{hev} & \harveyBallFull\\
\midrule
Feasible energy supply & \makecell[l]{Exclusion of \glspl{fcev}\\\SI{230}{\volt}\,\pdftooltip{\gls{ac}}{\glsentrydesc{ac}} charging\\Long battery range (\SI{>100}{\kilo\meter})} & \harveyBallFull\\
\midrule
\makecell[l]{Reliable operation\\Low maintenance requirements} & \makecell[l]{Simple design\\Enclosed components\\Proven components} & \harveyBallFull\\
\midrule
Easy maintenance & \makecell[l]{Use of standard parts\\Accessible modular construction\\Technical support available} & \harveyBallThreeQuarter\\
\midrule
Easy failure repair & \makecell[l]{Good spare parts supply\\Technical support available} & \harveyBallHalf\\
\midrule
Easy damage repair & \makecell[l]{Widely known materials\\Available joining processes\\Good spare parts supply} & \harveyBallThreeQuarter\\
\midrule
Uneven road surface & \makecell[l]{High ground clearance\\Long suspension travel\\Large ramp\,\&\,breakover angles} & \harveyBallFull\\
\midrule
\makecell[l]{Unpaved road surface\\Flooded or damaged roads} & \makecell[l]{All-wheel drive\\Off-road tires\\High wading depth} & \harveyBallThreeQuarter\\
\midrule
Inexperienced drivers & Easy-to-use cockpit layout & \harveyBallFull\\
\midrule
Inexperienced personnel & Safe-to-touch voltage levels (\SI{<60}{\volt}\,\pdftooltip{\gls{dc}}{\glsentrydesc{dc}}) & \harveyBallFull\\
\midrule
Versatile cargo space & Flat truck bed & \harveyBallFull\\
\midrule
Higher capacity than trad. \pdftooltip{\glspl{imt}}{\glsentrydesc{imt}} & min.\,\SI{1000}{\kilogram} useful load & \harveyBallFull\\
\midrule
Energy transport capability & External power outlet (\SI{230}{\volt}\,\pdftooltip{\gls{ac}}{\glsentrydesc{ac}}) & \harveyBallFull\\
\bottomrule
\end{tabular}
\end{table}

\subsection{Data Collection}
An optimized operational strategy and therefore vehicle distribution maximizes the impact electric vehicles can have in this project while at the same time reducing cost, energy consumption, and workload. To select an optimal operational strategy, simulations in mobility frameworks accurately depicting human behavior are needed. Conventionally, detailed mobility models are agent-based \cite{BernhardLuger.2017,Hager.2015} and need much input data for calibration \cite{Balmer.2008,Balmer.2009}. Therefore, innovative mobility models, methods, and data-acquisition devices shall be explored to reduce data requirements and to acquire needed data that describes the status quo of mobility in the target regions. Different technological options for mobility assessment were compared, considering the calibration requirements of existing mobility frameworks and on-site acquisition efforts. \autoref{tab:datacollmeth} shows the preferred methods that offer the best ratio between acquisition effort and data usability. The collected data will be shared openly to the scientific community and stakeholders after anonymization and visualized inside an online map platform for easy continuous apprehension and processing towards mobility framework inputs. This transformation process is based on participatory\,\&\,collaborative involvement by the open community and will be analyzed in further publications.
\begin{table}[!htbp]
\scriptsize
\centering
\caption{Data collection methods}
\label{tab:datacollmeth}
\begin{tabular}{lll}
\toprule
Source & Method & Data output\\
\toprule
GPS & OBD\,\&\,GPS logger & \makecell[l]{Vehicle position\\Altitude\\Speed\\Technical data}\\
\midrule
GPS & Smartphone app & \makecell[l]{Participant position\\Altitude\\Speed\\Modality\\Mobility behavior}\\
\midrule
Video & Traffic enumeration devices & \makecell[l]{Traffic flow\\Modalities\\Local speed}\\
\midrule
Survey &  Mobility survey & \makecell[l]{Mobility intentions\\Individual \pdftooltip{\glspl{poi}}{\glsentrydesc{poi}}}\\
\midrule
\makecell[l]{Observations\\Pictures\\Community engagement} & Map framework & \makecell[l]{Road infrastructure\\Electrical infrastructure\\Aggregated \pdftooltip{\glspl{poi}}{\glsentrydesc{poi}}}\\
\bottomrule
\end{tabular}
\end{table}
\subsection{Energy Sector Coupling}
In contrast to conventionally fueled vehicles, which rely on a separate infrastructure for their energy supply, \pdftooltip{\glspl{bev}}{\glsentrydesc{bev}} are inseparably linked to the local electric system. This effectively couples the sectors of electrification and motorization of transport and opens up many more research questions that were previously irrelevant. Some of those questions consider (1) the requirements that electric mobility poses to electric supply in a rural \pdftooltip{\gls{ssa}}{\glsentrydesc{ssa}} setting, (2) the assessment of these requirements in specific geographies, (3) the impact of \pdftooltip{\gls{ac}}{\glsentrydesc{ac}} power-sharing on agricultural use cases, (4) the implementation in rural settings, both centralized and decentralized and (5) the influence of all previous results on the optimal design of \pdftooltip{\glspl{bev}}{\glsentrydesc{bev}}.
These questions shall also be addressed within the project \quotes{aCar Mobility} through analysis of open-access data on electrification and gathered data from the implemented vehicle in conjunction with modelling\,\&\,mathematical optimization of power systems using open-source software. This enables toggling the presence of electric vehicles on or off, comparing optimum configurations in their cost structures, uncertainties and values offered to users. The focus will be on decentralized systems, as rural areas are their main application area \cite{BloombergNEF.2020}, they are facing scalability issues due to rarely being economically sustainable on their own \cite{Azimoh.2017,AllianceforRuralElectrification.2014,Pedersen.2016,Pachauri.2012}, and offer advantages in modeling due to being closed systems. Results shall be validated in existing electrification projects by transferring the test vehicle there temporarily. Should the need for modifications to the vehicle arise, prototypical installations are possible using the test vehicle both in Germany and Ethiopia.

The results of these efforts shall aid electrification and transport stakeholders such as legislators, utilities, developers and operators in their decision-making towards clean, affordable energy and mobility as well as  \pdftooltip{\gls{bev}}{\glsentrydesc{bev}} designers, producers and operators in more effectively targeting the rural \pdftooltip{\gls{ssa}}{\glsentrydesc{ssa}} market. All proceedings and results will be published with open-source code and open-access data wherever possible.
\section{Conclusion \& Discussion}
The research approach presented in this paper fills the gap between pure assessment methods and the actual implementation of technological solutions. Taking a pragmatic, result-oriented perspective on existing approaches, our work introduces a problem-to-solution procedure that combines a variety of methods to fulfill the contextual requirements. The overarching goal is a methodology that caters to research, development, and testing of vehicle-based solutions for agricultural supply and value chains in Ethiopia, ultimately focusing on the three \pdftooltip{\glspl{crp}}{\glsentrydesc{crp}} affordability, availability, and vehicle type. By encouraging user participation, early transfer of ownership, and continuous vehicle operation, we seek to address shortcomings of previous research projects and develop sustainable business models around novel technology.

In 2003, a report by the \pdftooltip{\gls{ssatp}}{\glsentrydesc{ssatp}} concluded that it is still easier to assess the ‘contribution’ of the transport sector in terms of deliverables, such as the number of kilometers of road rehabilitated… or the provision of \pdftooltip{\gls{imt}}{\glsentrydesc{imt}}, than in terms of improved access and mobility \cite{Braithwaite.2003}. Although this conclusion was drawn more than 17 years ago, measuring and improving people’s choices for transportation is still a pressing issue for most SSA countries. In particular, the perspective on farmers' ability to make crucial outsourcing decisions optimizing and specializing their supply and value chain introduced at the beginning of this paper is at stake. This inability has far-reaching consequences. Referring to Adam Smith’s theory of the division of labor, Rosenberg concludes his negotiation about the importance and risks of specialization regarding innovativeness with the remark that \quotes{the creativity of society as a whole grows while that of the laboring poor declines} \cite{Rosenberg.1965}. Increasing the yield per hectare is linked to a functioning supply chain and effective, value-creating farming activities, but there is an upper limit. Land is scarce, and continuous population growth in \pdftooltip{\gls{ssa}}{\glsentrydesc{ssa}} countries and in Ethiopia particularly increases the socio-economic pressure on individuals \cite{Abay.2020}. Rural smallholder farmers and their offspring need to seek diversification of income-generating activities off the farm \cite{AloboLoison.2015} which again puts affordable, available, and suitable transportation on the table.

The research approach introduced focuses on the agricultural supply and value chain as the system to be enhanced for the sake of impact measurability. Nevertheless, certain incorporated methods, like the living lab approach, enlarge the user’s degrees of freedom to utilize technology in more than only agriculture and in ways it was never used before. Consequently, from the implementation day, the proposed research approach is due to constant iteration and improvement. Believing in technology’s ability to enhance lives, we as engineers need to always be able to adapt any methodology to the needs and desires of people.
\section*{Acknowledgment}
As first author, C.\,P. devised the idea for this paper and created most of the content. P.\,R. contributed individual sections and conceptual as well as detailed revisions. D.\,Z. contributed individual sections of the paper. M.\,L. made an essential contribution to the conception of the research project. He critically revised the paper for its important intellectual content. M.L. gave final approval of the version to be published and agrees to all aspects of the work. As a guarantor, he accepts responsibility for the overall integrity of the paper. All authors have read and agreed to the published version of the manuscript.

This research was accomplished within and funded by the \quotes{aCar Mobility} research project in collaboration with \gls{giz} and funded by the \gls{bmz}. The authors declare no conflict of interest between funding and the presented research approach.

\ifCLASSOPTIONcaptionsoff
  \newpage
\fi



%\bibliographystyle{IEEEtran}splncs04

\bibliographystyle{IEEEtran}
\bibliography{literature}

\begin{thebibliography}{10}
\providecommand{\url}[1]{#1}
\csname url@samestyle\endcsname
\providecommand{\newblock}{\relax}
\providecommand{\bibinfo}[2]{#2}
\providecommand{\BIBentrySTDinterwordspacing}{\spaceskip=0pt\relax}
\providecommand{\BIBentryALTinterwordstretchfactor}{4}
\providecommand{\BIBentryALTinterwordspacing}{\spaceskip=\fontdimen2\font plus
\BIBentryALTinterwordstretchfactor\fontdimen3\font minus \fontdimen4\font\relax}
\providecommand{\BIBforeignlanguage}[2]{{%
\expandafter\ifx\csname l@#1\endcsname\relax
\typeout{** WARNING: IEEEtran.bst: No hyphenation pattern has been}%
\typeout{** loaded for the language `#1'. Using the pattern for}%
\typeout{** the default language instead.}%
\else
\language=\csname l@#1\endcsname
\fi
#2}}
\providecommand{\BIBdecl}{\relax}
\BIBdecl

\bibitem{Porter.2001}
M.~E. Porter, ``{The value chain and competitive advantage},'' in \emph{{Understanding business: processes}}, D.~Barnes, Ed.\hskip 1em plus 0.5em minus 0.4em\relax London: {Routledge in association with the Open University}, 2001, vol.~2, pp. 50--66.

\bibitem{Quinn.1994}
J.~B. Quinn and F.~G. Hilmer, ``{Strategic outsourcing},'' \emph{{MIT Sloan Management Review}}, vol.~35, no.~4, p.~43, 1994.

\bibitem{LaLonde.1994}
B.~J. {La Londe} and J.~M. Masters, ``{Emerging logistics strategies},'' \emph{{International Journal of Physical Distribution {\&} Logistics Management}}, vol.~24, no.~7, pp. 35--47, 1994.

\bibitem{Bardi.1991}
E.~J. Bardi and M.~Tracey, ``{Transportation outsourcing: a survey of US practices},'' \emph{{International Journal of Physical Distribution {\&} Logistics Management}}, vol.~21, no.~3, pp. 15--21, 1991.

\bibitem{Degu.2019}
A.~{Abebaw Degu}, ``{The Causal Linkage Between Agriculture, Industry and Service Sectors in Ethiopian Economy},'' \emph{{American Journal of Theoretical and Applied Business}}, vol.~5, no.~3, p.~59, 2019.

\bibitem{Abebaw.2013}
D.~Abebaw and M.~G. Haile, ``{The impact of cooperatives on agricultural technology adoption: Empirical evidence from Ethiopia},'' \emph{{Food Policy}}, vol.~38, pp. 82--91, 2013.

\bibitem{Abay.2020}
K.~Abay, K.~Hirvonen, and B.~Minten, ``{Farm Size, Food Security, and Welfare},'' in \emph{{Ethiopia's agrifood system: Past trends, present challenges, and future scenarios}}, P.~A. Dorosh and B.~Minten, Eds.\hskip 1em plus 0.5em minus 0.4em\relax Washington, DC: {International Food Policy Research Institute}, 2020, pp. 146--173.

\bibitem{Starkey.2016}
P.~Starkey, ``{Provision of rural transport services: user needs, practical constraints and policy issues},'' Bangkok, Thailand.

\bibitem{Wittenbrink.2014}
P.~Wittenbrink, \emph{{Transportmanagement: Kostenoptimierung, Green Logistics und Herausforderungen an der Schnittstelle Rampe}}, 2nd~ed.\hskip 1em plus 0.5em minus 0.4em\relax Wiesbaden: {Springer Gabler}, 2014.

\bibitem{Sieber.2009}
N.~Sieber, ``{Leapfrogging from Rural Hubs to New Markets: Rural Transport in Developing Countries},'' International Road Federation Bulletin, Special Edition Rural Transport.

\bibitem{Poulton.2010}
C.~Poulton, A.~Dorward, and J.~Kydd, ``{The Future of Small Farms: New Directions for Services, Institutions, and Intermediation},'' \emph{{World Development}}, vol.~38, no.~10, pp. 1413--1428, 2010.

\bibitem{Dennis.2017}
R.~Dennis and K.~Pullen, ``{Vehicles for rural transport services in sub-Saharan Africa},'' \emph{{Proceedings of the Institution of Civil Engineers - Transport}}, vol. 170, no.~6, pp. 321--327, 2017.

\bibitem{Ellis.1998}
\BIBentryALTinterwordspacing
S.~Ellis and J.~Hine, ``{The Provision of Rural Transport Services: Approach Paper}.'' [Online]. Available: \url{https://documents.worldbank.org/en/publication/documents-reports/documentdetail/691771468767412769/the-provision-of-rural-transport-services-approach-paper}
\BIBentrySTDinterwordspacing

\bibitem{Hine.2014}
J.~Hine, ``{Good Policies and Practices on Rural Transport in Africa: Planning Infrastructure {\&} Services},'' Washington, D.C.

\bibitem{Bryceson.2008}
D.~F. Bryceson, A.~Bradbury, and T.~Bradbury, ``{Roads to Poverty Reduction? Exploring Rural Roads' Impact on Mobility in Africa and Asia},'' \emph{{Development Policy Review}}, vol.~26, no.~4, pp. 459--482, 2008.

\bibitem{Njenga.2015}
P.~Njenga, S.~Willilo, and J.~Hine, ``{First Mile Transport Challenges for Smallholder Tomato Farmers along Ihimbo--Itimbo Road, Kilolo District Tanzania}.''

\bibitem{Njenga.2014}
P.~Njenga, G.~Wahome, and J.~Hine, ``{Pilot Study on First Mile Transport Challenges in the Onion Small Holder Sector}.''

\bibitem{Dennis.1998}
R.~Dennis, ``{Rural Transport and Accessibility: A Synthesis Paper},'' Geneva.

\bibitem{Sieber.2019}
N.~Sieber and P.~Njenga, ``{Modern Logistics for High Value Products in Sub-Saharan Africa},'' in \emph{{Urban and rural poverty}}, ser. {Hunger and poverty : causes, impacts and eradication}, D.~Oortwijn, Ed.\hskip 1em plus 0.5em minus 0.4em\relax New York: {Nova Science Publishers}, 2019.

\bibitem{Soltes.2018d}
M.~{\v{S}}olt{\'e}s, S.~Koberstaedt, M.~Lienkamp, S.~Rauchbart, and F.~Frenkler, ``{ACar-a Electric Vehicle Concept for Sub-Saharan Africa},'' in \emph{{2018 IEEE PES/IAS PowerAfrica}}, 2018, pp. 301--306.

\bibitem{Starkey.2019}
P.~Starkey, J.~Hine, R.~Workman, and A.~Otto, ``{Interactions between improved rural access infrastructure and transport services provision}.''

\bibitem{Starkey.2007b}
P.~Starkey, ``{Methodology for rapid assessment of rural transport services}.''

\bibitem{Donnges.2003}
C.~Donnges, \emph{{Improving Access in Rural Areas: Guidelines for Integrated Rural Accessibility Planning}}.\hskip 1em plus 0.5em minus 0.4em\relax Bangkok, Thailand: {International Labour Organization}, 2006.

\bibitem{Bryceson.2003}
D.~F. Bryceson, T.~C. Mbara, and D.~Maunder, ``{Livelihoods, daily mobility and poverty in sub-saharan Africa},'' \emph{{Transport Reviews}}, vol.~23, no.~2, pp. 177--196, 2003.

\bibitem{Banjo.2012}
G.~Banjo, H.~Gordon, and J.~Riverson, ``{Rural Transport: Improving its Contribution to Growth and Poverty Reduction in Sub-Saharan Africa}.''

\bibitem{Booth.2000}
D.~Booth, L.~Hanmer, and E.~Lovell, ``{Poverty and Transport: A report prepared for the World Bank in collaboration with DFID}.''

\bibitem{Minten.2020}
B.~Minten, M.~Dereje, F.~Bachewe, and S.~Tamru, ``{Evolving Food Value Chains},'' in \emph{{Ethiopia's agrifood system: Past trends, present challenges, and future scenarios}}, P.~A. Dorosh and B.~Minten, Eds.\hskip 1em plus 0.5em minus 0.4em\relax Washington, DC: {International Food Policy Research Institute}, 2020, pp. 177--219.

\bibitem{Creswell.2017}
J.~W. Creswell and J.~D. Creswell, \emph{{Research Design: Qualitative, Quantitative, and Mixed Methods Approaches}}, fifth edition~ed.\hskip 1em plus 0.5em minus 0.4em\relax Los Angeles and London and New Delhi: {Sage publications} and SAGE, 2017 // 2018.

\bibitem{Simon.1988}
H.~A. Simon, ``{The science of design: Creating the artificial},'' \emph{{Design Issues}}, vol.~4, no. 1/2, pp. 67--82, 1988.

\bibitem{Peffers.2007}
K.~Peffers, T.~Tuunanen, M.~A. Rothenberger, and S.~Chatterjee, ``{A Design Science Research Methodology for Information Systems Research},'' \emph{{Journal of management information systems}}, vol.~24, no.~3, pp. 45--77, 2007.

\bibitem{Meinel.2016}
C.~Meinel and J.~von Thienen, ``{Design Thinking},'' \emph{{Informatik-Spektrum}}, vol.~39, no.~4, pp. 310--314, 2016.

\bibitem{Chambers.1994}
R.~Chambers, ``{The origins and practice of participatory rural appraisal},'' \emph{{World Development}}, vol.~22, no.~7, pp. 953--969, 1994.

\bibitem{Eriksson.2006}
M.~Eriksson, V.-P. Niitamo, S.~Kulkki, and K.~A. Hribernik, ``{Living labs as a multi-contextual R{\&}D methodology},'' in \emph{{2006 IEEE International Technology Management Conference (ICE)}}, 2006, pp. 1--8.

\bibitem{Leminen.2012}
S.~Leminen, M.~Westerlund, and A.-G. Nystr{\"o}m, ``{Living Labs as open-innovation networks},'' \emph{{Technology Innovation Management Review}}, vol.~2, no.~9, pp. 6--11, 2012.

\bibitem{Vajjhala.2010}
S.~P. Vajjhala and W.~M. Walker, ``{Roads to Participatory Planning: Integrating Cognitive Mapping and GIS for Transport Prioritization in Rural Lesotho},'' \emph{{Journal of Maps}}, vol.~6, no.~1, pp. 488--504, 2010.

\bibitem{Hill.1999}
J.~Hill and P.~McGowan, ``{Small business and enterprise development: questions about research methodology},'' \emph{{International Journal of Entrepreneurial Behavior {\&} Research}}, 1999.

\bibitem{Mojo.2018}
D.~Mojo, T.~Degefa, and C.~Fischer, ``{The Development of Agricultural Cooperatives in Ethiopia: History and a Framework for Future Trajectory},'' \emph{{Ethiopian Journal of the Social Sciences and Humanities}}, vol.~13, no.~1, p.~49, 2018.

\bibitem{Getnet.2012}
K.~Getnet and T.~Anullo, ``{Agricultural Cooperatives and Rural Livelihoods: Evidence from Ethiopia},'' \emph{{Annals of Public and Cooperative Economics}}, vol.~83, no.~2, pp. 181--198, 2012.

\bibitem{EVUMMotorsGmbH.2019}
\BIBentryALTinterwordspacing
{EVUM Motors GmbH}, ``{aCar Press Images},'' 2019. [Online]. Available: \url{https://evum-motors.com/wp-content/uploads/2019/09/Studiobilder.zip}
\BIBentrySTDinterwordspacing

\bibitem{BernhardLuger.2017}
{Bernhard Luger}, ``{Generation of a Synthetic Population for MATSim Models Using Multidimensional Iterative Proportional Fitting and Discrete Choice Models},'' {Master's Thesis}, {TU Graz}, Graz, 2017.

\bibitem{Hager.2015}
K.~Hager, J.~Rauh, and W.~Rid, ``{Agent-based Modeling of Traffic Behavior in Growing Metropolitan Areas},'' \emph{{Transportation Research Procedia}}, vol.~10, pp. 306--315, 2015.

\bibitem{Balmer.2008}
M.~Balmer, K.~Meister, M.~Rieser, K.~Nagel, and K.~W. Axhausen, ``{Agent-based simulation of travel demand: Structure and computational performance of MATSim-T},'' 2008.

\bibitem{Balmer.2009}
M.~Balmer, M.~Rieser, K.~Meister, D.~Charypar, N.~Lefebvre, and K.~Nagel, ``{MATSim-T : Architecture and Simulation Times},'' 2009.

\bibitem{BloombergNEF.2020}
\BIBentryALTinterwordspacing
{Bloomberg NEF}, ``{State of the global Mini-grids Market Report 2020: Trends of renewable energy hybrid minigrids in Sub-Saharan Africa, Asia and island nations}.'' [Online]. Available: \url{https://minigrids.org/market-report-2020/}
\BIBentrySTDinterwordspacing

\bibitem{Azimoh.2017}
\BIBentryALTinterwordspacing
C.~L. Azimoh, P.~Klintenberg, C.~Mbohwa, and F.~Wallin, ``{Replicability and scalability of mini-grid solution to rural electrification programs in sub-Saharan Africa},'' \emph{{Renewable Energy}}, vol. 106, pp. 222--231, 2017. [Online]. Available: \url{www.scopus.com}
\BIBentrySTDinterwordspacing

\bibitem{AllianceforRuralElectrification.2014}
\BIBentryALTinterwordspacing
{Alliance for Rural Electrification}, ``{Hybrid Mini-Grids for Rural Electrification: Lessons Learned},'' Brussels. [Online]. Available: \url{https://www.ruralelec.org/publications/hybrid-mini-grids-rural-electrification-lessons-learned}
\BIBentrySTDinterwordspacing

\bibitem{Pedersen.2016}
\BIBentryALTinterwordspacing
M.~B. Pedersen, ``{Deconstructing the concept of renewable energy-based mini-grids for rural electrification in East Africa},'' \emph{{Wiley Interdisciplinary Reviews: Energy and Environment}}, vol.~5, no.~5, pp. 570--587, 2016. [Online]. Available: \url{www.scopus.com}
\BIBentrySTDinterwordspacing

\bibitem{Pachauri.2012}
S.~Pachauri, A.~Brew-Hammond, D.~F. Barnes, D.~H. Bouille, S.~Gitonga, V.~Modi, G.~Prasad, A.~Rath, H.~Zerriffi, T.~Dafrallah, C.~Heruela, F.~Kemausuor, R.~Kowsari, Y.~Nagai, K.~Rijal, M.~Takada, N.~Wamukonya, and J.~Sathaye, ``{Energy Access for Development},'' in \emph{{Global Energy Assessment (GEA)}}, A.~P. Patwardhan, L.~Gomez-Echeverri, N.~Naki{\'c}enovi{\'c}, and T.~B. Johansson, Eds.\hskip 1em plus 0.5em minus 0.4em\relax Cambridge: {Cambridge University Press}, 2012, pp. 1401--1458.

\bibitem{Braithwaite.2003}
M.~Braithwaite, ``{Report on the Pilot Country Case Studies of transport Policy and Poverty Reduction}.''

\bibitem{Rosenberg.1965}
N.~Rosenberg, ``{Adam Smith on the Division of Labour: Two Views or One?}'' \emph{{Economica}}, vol.~32, no. 126, pp. 127--139, 1965.

\bibitem{AloboLoison.2015}
S.~{Alobo Loison}, ``{Rural Livelihood Diversification in Sub-Saharan Africa: A Literature Review},'' \emph{{The Journal of Development Studies}}, vol.~51, no.~9, pp. 1125--1138, 2015.

\end{thebibliography}

%\begin{biography}[{\includegraphics[width=1in,height=1.25in,clip,keepaspectratio]{mshell}}]{Michael Shell}

\begin{IEEEbiographynophoto}{Clemens Pizzinini}
is team lead for electric vehicle system design at the Institute of Automotive Technology. Since 2020, his research focuses on the design of sustainable transportation systems in rural sub-Saharan Africa. In the aCar Mobility Research Project, he is responsible for the design and implementation of vehicle-based services. 
\end{IEEEbiographynophoto}

\begin{IEEEbiographynophoto}{Philipp Rosner}
is team lead for electric vehicle operations at the Institute of Automotive Technology. Since 2020, his research focuses on the integration of electric vehicles into mini grids, with a focus on rural sub-Saharan Africa. In the aCar Mobility Research Project, he is concerned with all questions regarding charging infrastructure. 
\end{IEEEbiographynophoto}

\begin{IEEEbiographynophoto}{David Ziegler}
is member of the research group for smart mobility at the Institute of Automotive Technology. Since 2020, his research focuses on mobility prediction models in rural sub-Saharan Africa. In the aCar Mobility Research Project, he is responsible for gathering primary mobility data. 
\end{IEEEbiographynophoto}

\begin{IEEEbiographynophoto}{Markus Lienkamp}
is the head of the Institute of Automotive Technology. His contributions across the whole value chain of electric vehicles is profound and well cited. Since 2014, he supervises the institute's activities in various African countries. 
\end{IEEEbiographynophoto}

%\newpage


%\vfill


\end{document}